\documentclass[prl,twocolumn,amssymb,amsmath,showpacs,superscriptaddress,groupedaddress]{revtex4}  

\usepackage[utf8]{inputenc} 
\usepackage{color}
\usepackage{colortbl}
\usepackage{graphicx}
\usepackage{amssymb,amsmath}
\usepackage{enumerate}
\usepackage{gensymb}
\usepackage{dsfont}
\usepackage[colorlinks=true,linkcolor=blue]{hyperref}
\begin{document}

\title[ATI]{Storage capacity and learning capability of quantum neural networks}

\author{Maciej Lewenstein}
\email[]{maciej.lewenstein@icfo.eu}
\affiliation{ICFO - Institut de Ci\`{e}ncies Fot\`{o}niques, The Barcelona Institute of Science and Technology, 08860 Castelldefels (Barcelona), Spain}
\affiliation{ICREA, Pg. Llu\'{\i}s Companys 23, 08010 Barcelona, Spain}

\author{Aikaterini Gratsea}
\affiliation{ICFO - Institut de Ci\`{e}ncies Fot\`{o}niques, The Barcelona Institute of Science and Technology, 08860 Castelldefels (Barcelona), Spain}

\author{Andreu Riera-Campeny}
\affiliation{F\'isica Te\`{o}rica: Informaci\'o i Fen\`{o}mens Qu\`{a}ntics.  Departament de F\'isica, Universitat Aut\`{o}noma de Barcelona, 08193 Bellaterra, Spain}

\author{Albert Aloy}
\affiliation{ICFO - Institut de Ci\`{e}ncies Fot\`{o}niques, The Barcelona Institute of Science and Technology, 08860 Castelldefels (Barcelona), Spain}

\author{Valentin Kasper}
\affiliation{ICFO - Institut de Ci\`{e}ncies Fot\`{o}niques, The Barcelona Institute of Science and Technology, 08860 Castelldefels (Barcelona), Spain}

\author{Anna Sanpera}
\affiliation{F\'isica Te\`{o}rica: Informaci\'o i Fen\`{o}mens Qu\`{a}ntics.  Departament de F\'isica, Universitat Aut\`{o}noma de Barcelona, 08193 Bellaterra, Spain}
\affiliation{ICREA, Pg. Llu\'{\i}s Companys 23, 08010 Barcelona, Spain}

\date{\today}

\begin{abstract}
We study the storage capacity of quantum neural networks (QNNs), described by completely positive trace preserving (CPTP) maps acting on a $N$-dimensional Hilbert space. 
We demonstrate that attractor QNNs can store in a non-trivial manner up to $N$ linearly independent pure states. For $n$ qubits, QNNs can reach an exponential storage capacity, $\mathcal O(2^{n})$, clearly outperforming classical neural networks whose storage capacity scales linearly with the number of neurons $n$. We estimate, employing the Gardner program, the relative volume of CPTP maps with $M\leq N$ stationary states and show that this volume decreases exponentially with $M$ and shrinks to zero for $M\geq N+1$. We generalize our results to QNNs storing mixed states as well as input-output relations for feed-forward QNNs. Our approach opens the path to relate storage properties of QNNs to the quantum features of the input-output states.  This paper is dedicated to the memory of Peter Wittek.  
\end{abstract}
\maketitle

Machine learning (ML), aiming at giving computers the ability to act without being explicitly programmed~\cite{Samuel1959}, crosses boundaries between such diverse fields as artificial intelligence, computer science, mathematics, physics, statistics, and neurosciences~\cite{Goodfellow2016}.  The roots of ML can be traced back to the last century, where the seminal concepts of artificial neurons and learning rules led to neural networks models (NNs), i.e., ensembles of interconnected neurons with learning capabilities. The present development of ML, largely based on the success of deep learning, has its roots in the 1980's when the statistical physics of NN was formulated and studied~\cite{Amit,youtube,Rumelhart}.


One crucial feature of NNs is their storage capacity for associative memory, that is, the number of {\it {patterns}} (stored memories/attractors) the network has for a given number of neurons $n$. For attractor NNs (aNNs) of the Hopfield-type~\cite{Hopfield}, where neurons are Ising spins and attractors correspond to metastable states resulting from spin-spin interactions, the relevant question is to determine how many stationary states, serving as stored memories, the network may have. For feed-forward NNs, with the paradigmatic example of the perceptron~\cite{Minsky}, the corresponding question is  how many attractor input-output relations can be stored. The problem of the storage capacity and learning ability of NNs was reformulated by the seminal contributions of Gardner~\cite{Gardner1,Gardner2}. She provided the relative volume of NNs with a desired set of patterns in the full space of NNs or, equivalently, the relative volume of feed-forward NNs with desired input-output relations. Sharp shrinking of the relative volume to zero, heralds the phase transition corresponding to an overloaded NN memory.


In the recent decades, quantum information science has demonstrated that information processing
can be significantly improved by exploiting quantum mechanics. Not surprisingly, both areas, ML and quantum information have merged together in the so-called quantum machine learning (QML)~\cite{Wittek,Biamonte}. QML encompasses different facets:  the application of ML techniques to quantum systems and devices \cite{Carleo,CarleoRMP}, and/or  the quantum implementation of ML concepts~\cite{Schuld2014,Deng2017}. 
The crucial question in this context is to determine whether QML offers quantum advantage, as demonstrated for instance in Noisy Intermediate-Scale Quantum (NISQ) devices \cite{Google}. A possible way to answer this question is to consider the storage capacity of QNNs.

Preliminary attempts to analyze the storage capacity of QNNs were pursued in~\cite{QPerceptron}. In a different approach, an exponential increase of the storage capacity for a specific quantum search algorithm was demonstrated in~\cite{Ventura98}. More recently~\cite{Rebentrost18, Meinhardt2020}, an increased storage capacity was obtained by using a feed-forward interpretation of quantum Hopfield NNs. Despite this progress, the storage capacity of generic QNNs remains an open problem.  In this work we address and solve this question by associating QNNs to CPTP maps.  We also analyze the learning capability of QNN's by applying Gardner's program to the quantum case and estimate the relative volume of QNNs realizing the desired attractor input-output relations.\\

\begin{figure}
    \centering
    \includegraphics[width=\columnwidth]{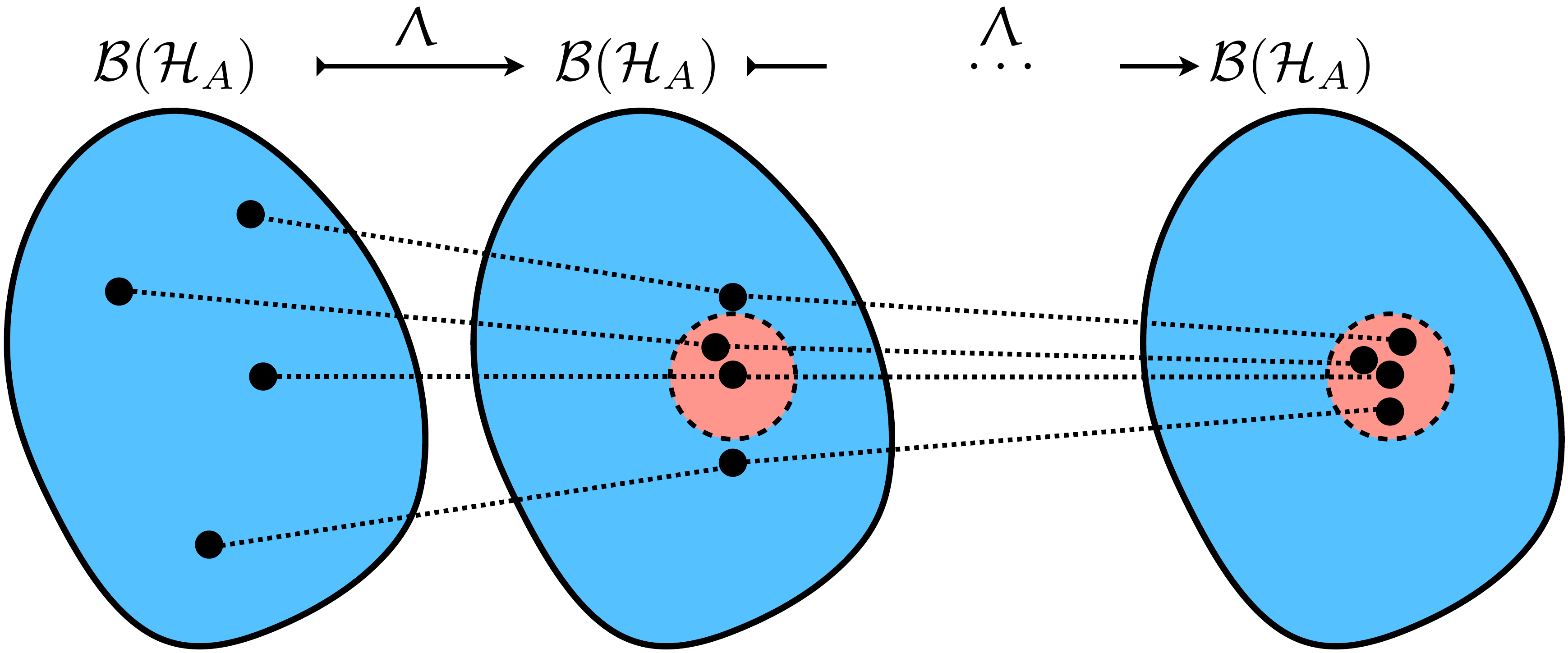}
    \caption{Color Online. Schematic representation of the action of  CPTP maps $\Lambda$ with $N$ fixed states.  Successive applications of  $\Lambda : \mathcal{B}(\mathcal{H}_A) \mapsto \mathcal{B}(\mathcal{H}_A)$, brings arbitrary states  $ \rho\in\mathcal{B}(\mathcal{H}_A)$ to the set (depicted by red area) of stationary states of the map.}
 \label{fig:1}
\end{figure}

 \noindent{\it Methods and results.} We associate QNNs with CPTP maps transforming initial
states into final states in a finite (or infinite) time. Attractors (stored memory/patterns) correspond to the stationary states of the map, i.e., $\Lambda(\rho)=\rho$. We identify the storage capacity of QNNs (number of stored memories) with the maximal number of stationary points of CPTP maps acting on density matrices in $N$-dimensional Hilbert spaces. We demonstrate that there exist a family of (non-trivial) CPTP maps that have $M=N$ linearly independent stationary pure states, and provide the generic expression of such maps. These maps act as attractors in the space of states, i.e., the successive application of the map brings an arbitrary state to the set of its fixed points, see Fig.~\ref{fig:1}. We interpret this class of maps as attractors QNNs (aQNNs). Further, we estimate the relative volume of CPTP maps that have exactly $1<M\leq N$ pure stationary states. This calculation corresponds to a quantum version of the Gardner program. We show that, in the limit of large $N$, the relative volume in the space of CPTP maps capable to store $M$ patterns decreases very slowly with $M$ as exp$(-M^2/(N^4-N^2))$. Our results clearly signal quantum advantage since CPTP maps acting on $n$-qubit states may reach a storage capacity of $\mathcal{O}(2^n)$. We derive analogous results for  bilayer QNNs and their respective attractor input-output relations. Finally, in the supplementary material (SM), we discuss the extension of our results to generic feed-forward NN.\\

\noindent{\it Storage capacity of attractive quantum neural networks.} 
To formalize the problem of the storage capacity of aQNN, we consider an input (output) Hilbert space $\mathcal{H}_A$ ($\mathcal{H}_B)$ of dimension $N_A$ ($N_B$), and denote by $\mathcal{B}(\mathcal{H}_A)$ ($\mathcal{B}(\mathcal{H}_B)$) their respective sets of linear bounded operators. 
Quantum states $\rho_A$ ($\rho_B$) correspond to positive semidefinite operators of unit trace in $\mathcal{B}(\mathcal{H}_A)$ ($\mathcal{B}(\mathcal{H}_B)$). Physical transformations can be characterized by CPTP maps, i.e., linear maps   $\Lambda : \mathcal{B}(\mathcal{H}_A) \mapsto \mathcal{B}(\mathcal{H}_B)$ fulfilling: (\textit{i}) positivity,  $\Lambda (\rho)\geq 0$, $\forall \rho\geq 0$; (\textit{ii}) complete positivity,  that is, any extension of the form $(\mathcal{I}\otimes\Lambda):\mathcal{B}(\mathcal{H}_C\otimes\mathcal{H}_A)\mapsto \mathcal{B}(\mathcal{H}_C\otimes\mathcal{H}_B)$ is also a positive map, where $\mathcal{I}$ is the identity map acting in an arbitrary space $\mathcal{H}_C$; and (\textit{iii}) trace preservation, $\mathrm{Tr}[\Lambda(\rho)] =\mathrm{Tr}[\rho]$.
A map $\Lambda$ can be characterized by an Hermitian operator $E_{\Lambda}\in{\cal B}({\cal H}_A\otimes {\cal H}_{B})$, via the Jamio\l kowski-Choi-Sudarshan (JCS) isomorphism~\cite{Jamiolkowski,Choi, Sudarshan}.  For  CPTP maps, the corresponding JCS operator is positive semidefinite $E_\Lambda \geq 0$ and fulfills (trace preserving condition) $\mathrm{Tr}_B[E_\Lambda]=\mathds{1}_A$. The JCS operator reads  
$E_\Lambda = (\mathcal{I}\otimes\Lambda)(|\Omega\rangle\langle \Omega|)$, where $|\Omega\rangle = \sum_{i=1}^{N_A} |i\rangle |i\rangle$ is an unnormalized maximally entangled state in $\mathcal{H}_A\otimes\mathcal{H}_{A'}$, with $\mathcal{H}_{A'}$ a duplicate of the input space. In fact, this constitutes an isomorphism since $\Lambda(\rho_A) = \mathrm{Tr}_A[E_\Lambda\rho^T_A]$.

It is is well known that each CPTP map has, at least, one stationary state. Here we investigate which is the maximal number of linearly independent stationary states that non-trivial maps ($\Lambda\ne {\cal I}$ ) may have. In what follows, we address first this question assuming that stationary states correspond to projectors onto pure states. Unless specified, henceforth we take the Hilbert spaces dimensions to be $N_A=N_B=N$.\\

\noindent{\bf Theorem 1} There exist non-trivial CPTP maps $\Lambda$ s.t. $\Lambda(|r_\mu\rangle\langle r_\mu|) = |r_\mu\rangle\langle r_\mu|$, where  $\{|r_\mu\rangle\}$ are linearly independent and $\mu = 1,\ldots, N$. 
 \label{eq:teorem1}\\

\noindent{\it Proof:} 1) First, we transform $\Lambda$ into a canonical form by noting that there always exist a linear, invertible transformation $T$, s.t. $|r_\mu\rangle =T|\mu\rangle$, where $\{|\mu\rangle\}_{\mu=1}^N$ form an orthonormal basis. Such transformation is unique up to :  (a) the choice of the basis; (b) the phases of the basis elements that cancel in the projectors; and (c) the permutations of the elements of the basis.  We define the canonical form $\tilde{\Lambda}$ as  $\tilde\Lambda(\rho)=T^{-1}\Lambda(T\rho T^\dag)(T^\dag)^{-1}$, which has the property that if 
$\tilde\Lambda( |\mu\rangle\langle\mu|)=|\mu\rangle\langle\mu|$ then
 $\Lambda(|r_\mu\rangle\langle r_\mu|) = |r_\mu\rangle\langle r_\mu|$.\\
  2) We decompose the corresponding JCS operator as $E_{\tilde{\Lambda}}= E_\mathcal{I} + Q\ge 0$, where $E_\mathcal{I}$ is the JCS operator corresponding to the identity map. Then, non-triviality of $E_{\tilde{\Lambda}}$ requires that $Q\ne 0$. By definition $Q= Q^\dag$, and ${\rm Tr}_B [Q]=0$ (trace preserving). Moreover, for every $|\mu\rangle$ we have that $\langle \mu| Q |\mu\rangle=0$, ergo $\langle \nu|\langle \mu| Q |\mu\rangle|\nu\rangle=0$ for any $\mu$, $\nu$. This implies that $\langle \mu \nu|E_{\tilde{\Lambda}}|\mu\nu\rangle$=0 for $\mu\ne \nu$ and, because  $E_{\tilde{\Lambda}}\ge 0$, then $E_{\tilde{\Lambda}} |\mu\nu\rangle=0$ for $\nu\ne \mu$. As a consequence, $Q |\mu\nu\rangle=0$ for $\mu\ne \nu$, which implies that $Q$ has only nonzero matrix elements in the subspace spanned by the vectors $|\mu \mu\rangle$.  Therefore, $Q=\sum_{\mu\nu}\alpha_{\mu\nu}|\mu\mu\rangle\langle\nu\nu|$, with $\alpha_{\mu \mu}=0$.
Finally, from $E_{\tilde{\Lambda}}\ge 0$, it follows that $|1+\alpha_{\mu\nu}|^2\le 1$ for all $\mu\neq \nu$. \\

Such maps cause reduction of coherences in the orthonormal basis $\{\vert{\mu}\rangle\}$, namely  if $\rho'=\Lambda(\rho)$, we find $|\langle\mu|\rho'|\nu\rangle| = |1+\alpha_{\mu\nu}||\langle\mu|\rho|\nu\rangle| \leq |\langle\mu|\rho|\nu\rangle|$ for $\mu\neq\nu$. The multiple iteration of these maps lead generically to a total decay of coherences (if for all $\mu\neq \nu$, $|1+\alpha_{\mu\nu}|<1$), and the memories stored will correspond to the fixed points of the dynamics (see Fig.\ref{fig:1}). Notice also that  if $|1+\alpha_{\mu\nu}|=1$ for $\mu\neq \nu$, the state $\vert\varphi\rangle= a\vert\mu\rangle + b\vert\nu\rangle$ is also stationary. From the above theorem we immediately obtain:\\

\noindent{\bf Lemma 1} Any CPTP map $\Lambda$, s.t. $\Lambda(|\mu\rangle\langle\mu|) = |\mu\rangle\langle \mu|$, where $\{|\mu\rangle \}_{\mu=1}^N$ forms an orthonormal basis, 
has associated a JCS operator of the form:
\begin{align}
     E_{\Lambda} = \sum_{\mu}^N |\mu\mu\rangle\langle\mu\mu| +\sum_{\mu\neq \nu}^{N} (1+\alpha_{\mu\nu})|\mu\mu\rangle\langle\nu\nu|,
 \label{eq:eea}
\end{align}
with $\alpha_{\mu\nu} \in \mathbb{C}$ and $|1+\alpha_{\mu\nu}|\leq 1$. \\

\noindent{\bf Corollary 1} Since a non-trivial $\Lambda$ exists for $M=N$, there are even more such maps for $M<N$.\\

\noindent{\bf Corollary 2} If $M\ge N+1$, that is, $\Lambda( |\mu\rangle\langle\mu|)=|\mu\rangle\langle\mu|$  for $\mu=1,\cdots,N$ and $\Lambda( |e\rangle\langle e |)=|e\rangle\langle e |$, where $\vert e\rangle=\sum_{\mu=1}^{N} c_{\mu}\vert\mu\rangle$ with all $c_{\mu}\neq 0$, then the map is trivial, $\Lambda\equiv {\cal I}$.\\

This implies that the non-trivial CPTP maps with $M\geq N+1$ stationary pure states ceases to exist if for all $\mu\neq\nu$, $\vert 1+\alpha_{\mu,\nu}\vert <1$, which is, generically, the case.\\

\begin{figure}
    \centering
    \includegraphics[width=0.35\textwidth]{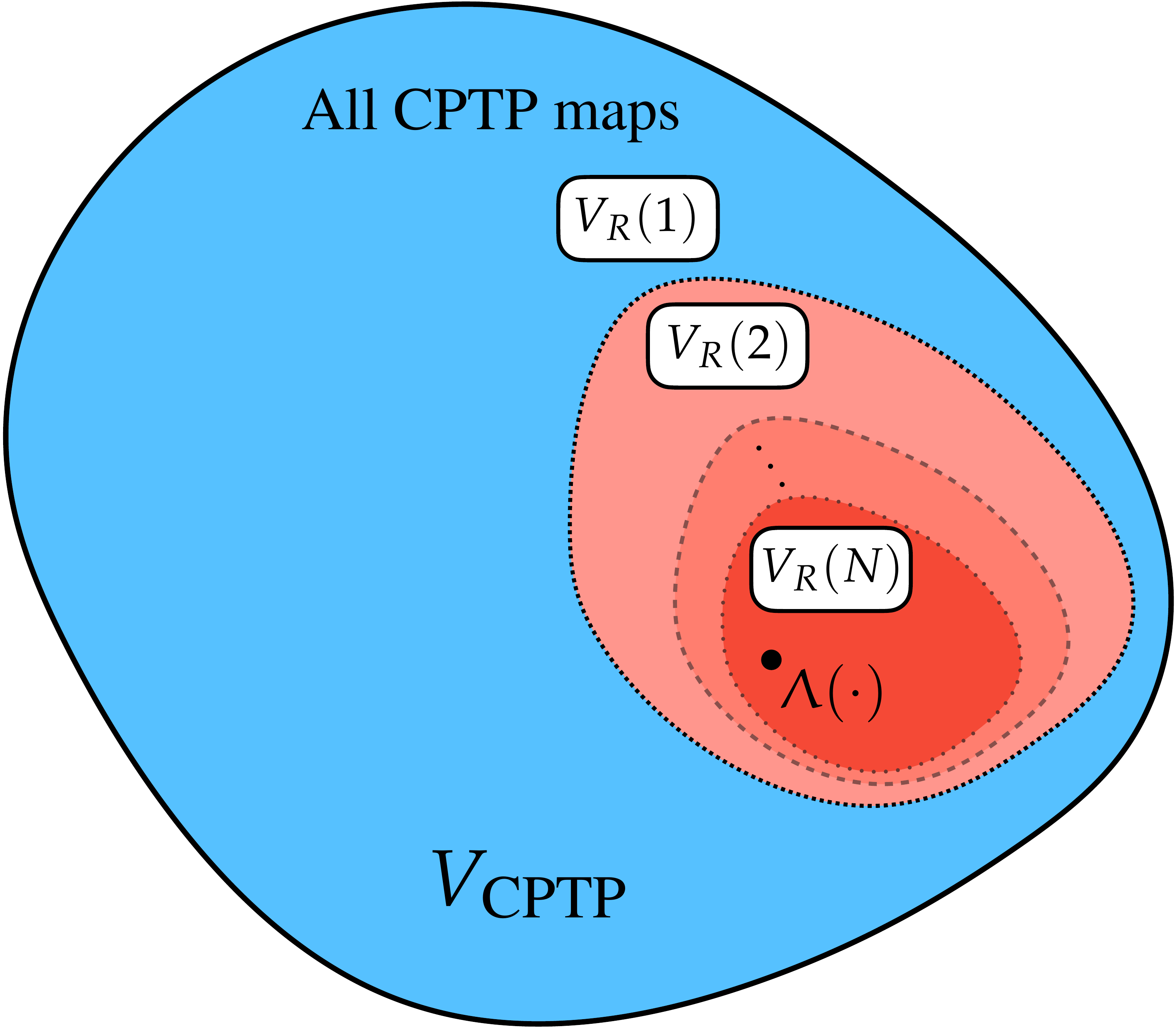}
    \caption{Color online.  Representation of the relative volume $V_R(M)$ of CPTP maps acting as aQNN and storing $M$ stationary pure states. The volume shrinks as we increase the number of stationary states from $V_\text{CPTP}=V_{R}(1)$ for $M=1$,  to $V_{R}(N)$ for $M=N$.}
      \label{fig:2}
\end{figure}

\noindent{\it Quantum Gardner program.}  The total volume of aQNN corresponds to the volume of CPTP maps that have a fixed set of stationary states. The volumes of various sets of maps have been estimated in \cite{Cappelini07, Szarek08} using the JCS isomorphism. The approach used by Szarek {\it et al.}  \cite{Szarek08, Dvoretzky} estimates, using the Hilbert-Schmidt norm, the radius of a ball that approximates the volume of CPTP maps in the asymptotic limit. In this limit, $N\to \infty$, this radius is $R=\exp(-1/4)$. The manifold of CPTP maps acting on a Hilbert space of dimension $N$, has dimension $d=N^4-N^2$, which corresponds to the dimension of the space of Hermitian JCS matrices ($N^4$), minus the number of real constraints imposed by trace-preserving condition ($N^2$). Using the volume of the unit ball, in the limit of sufficiently large $N$,  the volume of the CPTP manifold approximates~\cite{Szarek08} to:
\begin{equation}
V_{\rm CPTP} (d) =\frac{\pi^{d/2}}{\Gamma(d/2 + 1)}\exp(-d/4).
\label{eq_vcptp}
\end{equation}
An estimate for the volume of the aQNN manifold with exactly $M$ stationary linearly independent pure states, is obtained by imposing $M$ conditions in the $d$-dimensional space of the CPTP manifold: 
\begin{equation}
V_{\rm aQNN}(\epsilon,M,d)=\int d^{d} V_{\text{HS}} \prod_{\mu=1}^M \mathbf{1}_{[1-\epsilon/2,1+\epsilon/2]} \left(\langle \mu \mu| E_\Lambda |\mu\mu\rangle \right),
\label{eq_vaqnn}
\end{equation}
where $d^dV_{\text{HS}} $ is the Hilbert-Schmidt measure~\cite{Zyczkowski_2003} and $\mathbf{1}_{[1-\epsilon/2,1+\epsilon/2]}(x)$ is the indicator function being one for $x\in [1-\epsilon/2,1+\epsilon/2]$ and zero otherwise. The parameter $\epsilon$ defines a basin of attraction.  By definition, $V_{\rm aQNN}$ must be smaller than $V_{\rm CPTP}$. For  sufficiently small $\epsilon$, we may approximate $\mathbf{1}_{[1-\epsilon/2,1+\epsilon/2]}(x)\simeq \epsilon\delta(x-1)$. In this case, the integral over the $d$-dimensional manifold of CPTP maps with the $M$ constraints reduces from
 $d$ to $d-M$ dimensions. For $1\ll M\ll N$ but still large $d$, the radius of the corresponding ball 
remains asymptotically the same, and the volume of CPTP maps with $M$ stationary states (see Fig.~\ref{fig:2}) becomes 
\begin{equation}
V_{\rm aQNN}(\epsilon,M,d)\simeq \frac{\epsilon^M\pi^{(d-M)/2}}{\Gamma((d-M)/2 + 1)}e^{-(d-M)/4}.
\label{eq_vann2}
\end{equation}
Note that  this result does not depend on the concrete choice of the stationary states $\{|\mu \mu \rangle\}$.  Notice  also that, as shown in Lemma 1, there are infinitely many  CPTP maps with $M$ stationary states, but their volume is of measure zero for $\epsilon=0$.\\
The relative volume reads then: 
\begin{equation}
V_{R}(\epsilon,M,d)=  \frac{V_{\rm aQNN}(\epsilon,M,d)}{V_{\rm CPTP}(d)}\simeq \frac{\epsilon^{M} e^{M/4}\pi^{-M/2} (d/2)!}
{((d-M)/2)!} .
\label{eq_rvaqnn}
\end{equation}
Using Stirling's formula we obtain: 
\begin{align}
\ln V_R(\epsilon,M,d) \simeq  \frac{M}{2}\ln \left({\frac{ \sqrt{e} d \epsilon^2}{2\pi}}\right)-\frac{M^2}{4d}.
\end{align}
The choice of the parameter $\epsilon$ should be sufficiently small in order to be consistent with $V_R(\epsilon,M,d)<1$, which follows from Eq.~\eqref{eq_vaqnn}. Since we are interested in the scaling with $M$, an upper bound corresponds to setting $\epsilon=e^{-1/4}\sqrt{(2\pi)/d}$. Then the relative volume scales as
\begin{equation} 
V_R(M,d) \simeq \exp(-M^2/4d),\label{eq:vr_optimal}
\end{equation}
shrinking surprisingly slowly with $M$. Hence, the learning of $M\ll N$ patterns should be feasible for aQNNs. In particular, for systems of $n$ qubits where $N=2^n$,  $M$ can be of order $2^{n/2}$, that is, exponential in the number of qubits.\\
In what follows, we generalize our results to the case where the fixed points (stored memories) correspond to mixed states. To this aim we introduce the so-called {\it classical ensembles} as defined recently by Kronberg \cite{Kronberg}.\\

\noindent{\bf Definition 1} Let ${\cal E} = \{\rho_\mu \}$ with $\mu=1,\cdots,M$ be an ensemble of $N$-dimensional density matrices in $\mathcal{B}(\mathbb C^{N})$. The ensemble ${\cal E}$ is called \textit{classical} if there exists a single invertible operation $T$ that diagonalizes all elements of the ensemble; i.e., $T\rho_\mu T^\dag=D_\mu$, where all $D_\mu$ are simultaneously diagonal. We call this basis the {\it computational basis}.\\
The above definition generalizes the one given in~\cite{Kronberg}, since in our case $T$ does not have to be unitary. Although the maximal number of linearly independent density matrices in ${\cal E}$ equals $N$, the ensemble may contain many more elements, $M\ge N$. 

  
\noindent{\bf Theorem 1'} There exist non-trivial CPTP maps $\Lambda$, s.t. $\Lambda(\rho_{\mu}) = \rho_{\mu}$, where  $\rho_{\mu}\in {\cal E}$ with $\mu = 1,\ldots, M$, and arbitrary $M$.\\

\noindent{\it Proof:} The ensemble is determined by the complete basis in which all elements are diagonal. The required map, up to the canonical transformation to the corresponding orthonormal basis $\{|\mu\rangle\}$ , has the form given by Eq. \eqref{eq:eea}.  Since $\Lambda (|\mu\rangle\langle\mu|)=|\mu\rangle\langle\mu|$ for all $|\mu\rangle\langle\mu|$, then 
$\Lambda (\rho_{\mu})=\rho_{\mu}$ is also true for any $\rho_{\mu}$ that is a mixture of projectors $|\mu\rangle\langle\mu|$ and, therefore, for all the members of ${\cal E}$. 

The relative volume of the aQNN for the ensemble ${\cal E}$ behaves, however, differently that in the case of storing linearly indepedent pure states. Now, having $M\leq N$ stationary mixed states $\Lambda (\rho_{\mu})=\rho_{\mu}$ for $\mu=1,\cdots,M$, demands imposing $\sim M\times N^{2}$ constrains in the $d$-dimensional space of the CPTP manifold. In turn, this means that  the relative volume of CPTP maps storing $M$ mixed states should behave approximately as  $V_{R} (M,d)\sim e^{-M^{2}}$, decreasing very rapidly with $M$.\\

\noindent{\it Storage capacity of feed-forward QNN.} The generalization of the above results to feed-forward QNNs is presented in the Supplementary Material. There, we consider the case corresponding to different input and output dimensions. \\

 \noindent{\it Conclusions and Outlook.}  We have demonstrated, using CPTP maps acting on a Hilbert space of dimension $N$, that aQNN's can store up to $N$ linear independent pure states. For $n$ qubits, quantum channels reach thus the capacity $2^n$, clearly outperforming  the storage capacity of classical neural networks $\sim O(n)$, where $n$ is the number of binary neurons. Applying Gardner's program to the quantum case, we have related the learning capability of aQNN's  to the relative volume $V_{R}(M)$ of CPTP maps with $M$ stationary pure states, and show that this volume decreases very slowly with the number of stored patterns $M$. Finally, we have applied our procedure also to feed-forward QNN with different input and output spaces. Our results are simple and mathematically rigorous. Furthermore, they open the path to study the relation between the storage capacity of QNNs and the quantum features, such as coherence and entanglement, of the desired attractor input-output relations. \\

\acknowledgments{We acknowledge financial support from: ERC-AdG NOQIA, 
Spanish MINECO: FIS2016-79508-P,  FIS2016-80681-P (AEI/FEDER, UE),``Severo Ochoa'' program for Centers of Excellence in R\&D (CEX2019-000910-S),  Spanish Agencia Estatal de Investigaci\'on: (PID2019-107609GB100, PID2019-106901GB-I00 /10.13039 / 501100011033, FPI), European Social Fund, Generalitat de Catalunya: (CIRIT 2017-SGR-1341,2017-SGR-1127, AGAUR FI-2018-B01134, CERCA Program, and  QuantumCAT/001-P-001644, QuantumCAT\_U16-011424 co-funded by ERDF Operational Program of Catalonia 2014-2020), Fundaci\'o Privada Cellex, Fundaci\'{o} Mir-Puig,  MINCIN-EU QuantERA  MAQS funded by the State Research Agency (AEI): (PCI2019-111828-2, 10.13039/501100011033).
This project has received funding from the European Union Horizon 2020: PROBIST 754510, Marie Sk\l odowska-Curie grant agreement No. 754510, FET-OPEN OPTOLogic No 899794, and the National Science Centre Poland-Symfonia Grant No. 2016/20/W/ST4/00314.

\bibliography{references}

\appendix

\section{Supplementary material}

\noindent{\it Storage capacity of feed-forward QNN.} Feed-forward neural networks classify inputs states according to output states and the relevant question is how many input-output relations can be realized. This scenario corresponds to CPTP maps $\Lambda : \mathcal{B}(\mathcal{H}_A) \to \mathcal{B}(\mathcal{H}_B)$ such that $N_A\neq N_B$. We denote by $\rho$ a general input state and by $\sigma$ a given output state. In this case, a stored memory corresponds to an input-output relation of the form $\Lambda(\rho_{\mu(k)}) = \sigma_{k}$, for $\mu=1,\cdots, M_k$. The simplest example of a feed-forward QNN is the quantum perceptron which classifies any input state into two linearly independent output states with $k=0,1$  corresponding to $N_{A}=N$ and $N_{B}=2$ (see Fig.~\ref{fig:3}). For simplicity, we consider both the input and output states to be linearly independent pure states. In such case, the following theorem follows:

\begin{figure}
    \centering
    \includegraphics[width=\columnwidth]{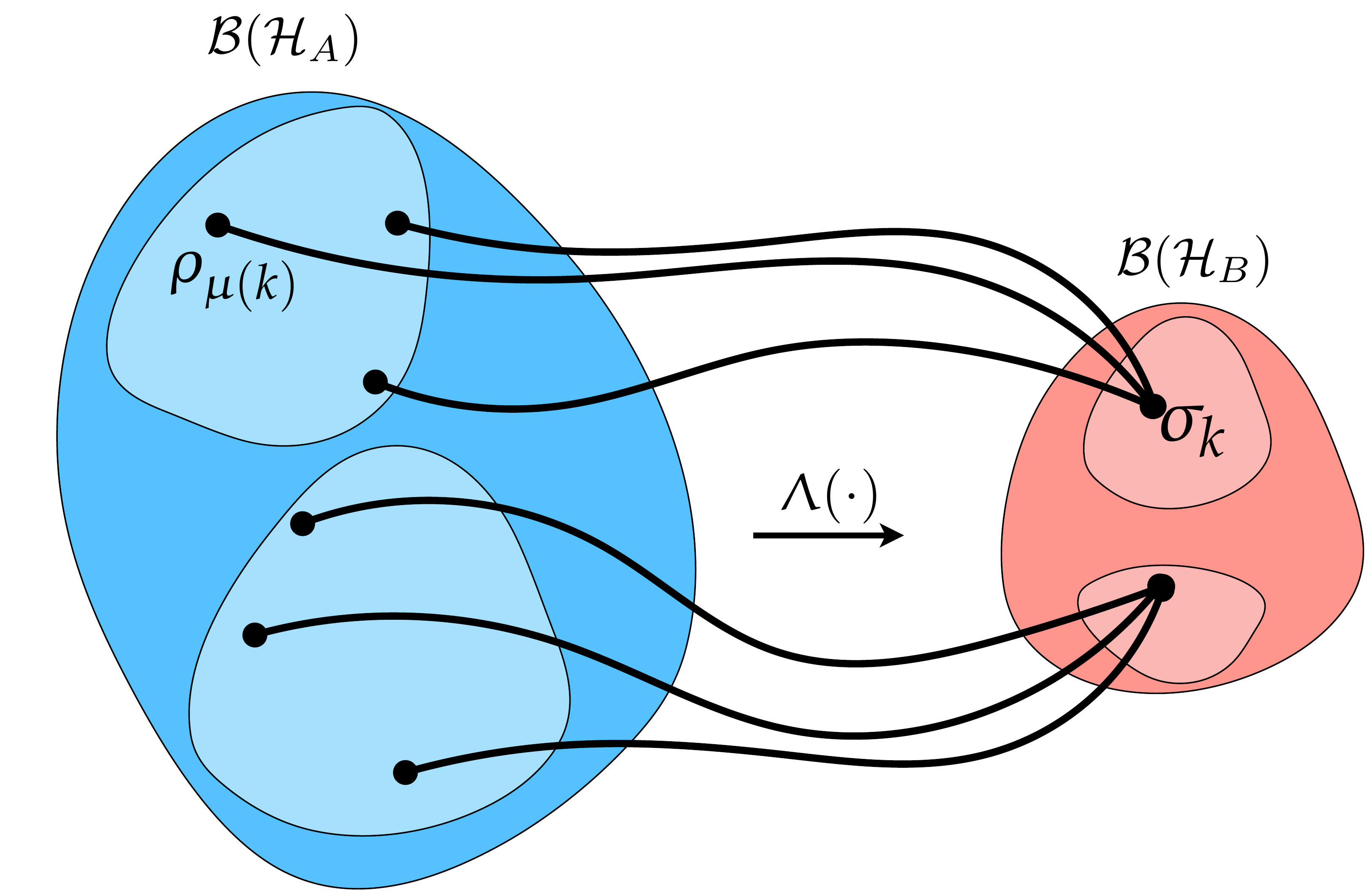}
    \caption{ Color online. Schematic representation of aQNNs associated to CPTP maps that classify input states, $\rho_\text{in} \in \mathcal{B}(\mathcal{H}_A)$, into two output states $ \rho_\text{out} \in  \mathcal{B}(\mathcal{H}_B)$.}
    \label{fig:3}
\end{figure}
\noindent{\bf Theorem 2} There exist a family of non-trivial CPTP maps $\Lambda$, s.t. 
$\Lambda(|r_{\mu}\rangle\langle r_{\mu}|) = |0\rangle\langle0|$ for $\mu=1, \ldots, M_0$,   
and $\Lambda(|r_{\mu}\rangle\langle r_{\mu}|) = |1\rangle\langle1|$ for $\mu=M_{0}+1,\dots N$, 
where $\{ |r_{\mu}\rangle \}$  and $(\{ |0\rangle, |1\rangle\})$ are linearly independent pure states in $ \mathcal{H}_A$ and $\mathcal{H}_B$ respectively.\\

\noindent{\it Proof:} The proof is analogous to the proof of Theorem 1 and follows the same steps. 
First, we transform $\Lambda$ into its canonical form, which now requires two linear transformations $T_A$ and $T_B$. Namely, there exists a linear, invertible transformation $T_A$, such that $|r_{\mu}\rangle =T_A|\mu\rangle$, where $\{|\mu\rangle\}$ for $\mu=1, \ldots, N$ forms an orthonormal basis in $\mathcal{H}_A$. Similarly, there exists a linear, invertible transformation $T_B$ such that $|s_{k}\rangle = T_{B}|k\rangle$, where $ \{|k\rangle\}$ for $k=0,1$ is an orthonormal basis in $\mathcal{H}_B$. We then define the canonical form $\tilde \Lambda(\rho)=T_B^{-1}\Lambda(T_A\rho T_A^\dag)(T_B^\dag)^{-1}$ so that 
$\tilde \Lambda(|\mu\rangle \langle {\mu}|)
=\vert{0}\rangle\langle 0\vert$ for $\mu=1,\dots M_{0}$ and 
$\tilde \Lambda(|\mu\rangle \langle {\mu}|)=\vert{1}\rangle\langle 1\vert)$ for $\mu=M_{0}+1,\dots N$. To make the notation more compact, we denote by $\{ \vert \mu_{0}\rangle \}$ with $\mu_{0}=1,\dots,M_{0}$, the input states whose output state is  $\vert 0 \rangle \langle 0\vert$, and  by
  $\{ \vert \mu_{1}\rangle \}$ for $\mu_{1}=1,\dots,M_{1}$, the input states whose output states is 
  $\vert 1 \rangle \langle 1\vert$. Of course $M_{0}+M_{1}=N$. Then we write the corresponding JCS operator with   
%
%
 $E_{\tilde{\Lambda}}= \sum_{\mu} |\mu_{k} k\rangle\langle\mu_{k} k | + Q\ge 0$, where the first term in the decomposition implements ``trivially'' the input-output relation. Non-triviality of $E_{\tilde{\Lambda}}$ requires that $Q \ne 0$. As before, $Q= Q^\dag$, and ${\rm Tr}_B [Q]=0$ (trace-preserving condition), except now the output dimension $N_B=2$. Moreover, $\langle \mu_{k}| Q |\mu_{k}\rangle=0$ for $\mu_{k}=1,\dots M_k$ and $k=0,1$.  Similarly as before, $\langle \mu_{k} k'|E_{\tilde{\Lambda}} |\mu_{k} k'\rangle = \delta_{kk'}$. The complete positivity of the map implies $E_{\tilde{\Lambda}} \ge 0$, therefore we have 
$E_{\tilde{\Lambda}}|\mu_{k} k'\rangle=0$ for $k\neq k'$. Therefore, we arrive at the full charaterisation $E_{\tilde{\Lambda}} = \sum_\mu |\mu_k k\rangle\langle \mu_{k} k| + X \otimes |0\rangle\langle 1| + X^\dagger \otimes|1\rangle\langle0|$ where $X$ is an arbitrary $M_0 \times M_1$ matrix. The positive definiteness of $E_{\tilde{\Lambda}}$ forces that $XX^\dag \le {\mathds 1_{M_0}}$ (or equivalently  $X^\dag X \le {\mathds 1_{M_1}}$), which means that all their eigenvalues must be smaller or equal than one. Finally, notice that the matrices $X$ and $X^\dag$ cause also the reduction of coherences of $\rho$ in the computational basis.\\

 The generalization of the above results to feed-forward QNN with more than two outputs, that is,  $\Lambda: {\cal B}(\mathcal{H}_A) \to {\cal B}(\mathcal{H}_B)$, with $N_B=3,4,5, \ldots$ is more technical but straightforward. Also its extension to a multi-layer QNN can be obtained by concatenating several maps. Similarly to the aQNN case, the Theorem 2 can be generalized the input states forming a {\it classical ensemble} \`a la Kronberg \cite{Kronberg}. 
 
We conclude by discussing the possibility of obtaining non-trivial CPTP maps able to implement a quantum perceptron that categorizes, for instance, $N^{2}$ pure input states to $N^2$ mixed output states. To this aim, we consider a composite input system ${\cal H}_A\otimes{\cal H}_{A'}$ where $\dim(\mathcal{H}_A)=\dim(\mathcal{H}_{A'})=N$. It can be easily shown that:\\

\noindent{\bf Lemma 2} There exists a basis $\{|\Psi^{AA'}_{\mu}\rangle\}\in{\cal H}_A\otimes{\cal H}_{A'}$, with $\mu=1, \ldots, N^2$, s.t. $\rho^A_\mu={\rm Tr}_{A'}[|\Psi^{AA'}_{\mu}\rangle \langle \Psi^{AA'}_{\mu}|]$ for $\mu=1, \ldots, N^2$ form a basis in the space of operators ${\cal B}({\cal H}_A)$.

\noindent{\it Proof:}  Any random set of $N^2$, in general, entangled states $\{|\Psi^{AA'}_{\mu}\rangle\}$, provides such set with probability arbitrarily close to one. Alternatively, we can also realize this task using $N^2$ product vectors  $|r_\mu\rangle\otimes |r'_\mu\rangle$, s.t. $|r_\mu\rangle\ne |r_\nu\rangle$ and $|r'_\mu\rangle\ne |r'_\nu\rangle$ for all $\mu \ne \nu$. The resulting basis in ${\cal B}({\cal H}_A)$ consists, in this case of projectors onto $N^2$ different pure states. Yet another constructive way to choose a desired set of $N^2$ linearly independent reduced density matrices $\rho^A_\mu$, is to diagonalize them and purify them into ${\cal H}_A\otimes{\cal H}_{A'}$, using random unitary bases in ${\cal H}_{A'}$.\\ 

\noindent A consequence of the Lemma 2, for $N_A = N_{A'} = N_B = N$ it follows:\\

\noindent{\bf Theorem 3}  There exist a family of non-trivial CPTP maps $\Lambda$ s.t.
 $\Lambda(|\Psi^{AA'}_{\mu}\rangle \langle \Psi^{AA'}_{\mu}|) =  \rho^B_\mu$, where $\{|\Psi^{AA'}_{\mu}\rangle\} ,  \ \mu=1, \ldots, N^2$ form a basis in ${\cal H}_A\otimes{\cal H}_{A'}$,  whereas $\{\rho^B_\mu\},  \ \mu=1, \ldots, N^2$ forms an operator basis in ${\cal B}({\cal H}_B)$.\\

 \noindent{\it Proof:} It follows from Lemma 2 that the basis set $\{|\Psi^{AA'}_{\mu}\rangle\} ,  \ \mu=1, \ldots, N^2$ exists. From Theorem 1 we know there exist a family of CPTP maps that have the pure states $\{|\Psi^{AA'}_{\mu}\rangle\}$ as fixed points. Hence, concatenating any of those maps with tracing over $A'$ one obtains the desired map $\Lambda(\cdot): {\cal B}({\cal H}_A\otimes{\cal H}_{A'} ) \to {\cal B}({\cal H}_B)$ such that $\Lambda(|\Psi^{AA'}_{\mu}\rangle \langle \Psi^{AA'}_{\mu}|) =  \rho^B_\mu$.\\
 
 This map is an extremal example and can be straightforwardly generalized to the case where $N_A = N_B = N$ and $N_{A'}=N'$ to obtain a generalized quantum perceptron capable of categorizing $NN'$ inputs ($N'>N$) into $N^2$ categories. \\
 
\end{document}